\documentclass[aps,pre,groupedaddress,floatfix]{revtex4}
\usepackage{graphicx}
\usepackage{amsmath}

\begin{document}
%\pagestyle{plain}
%\thispagestyle{empty}

%\mbox{} \vspace{1cm}

\title{Brownian motion: a case of temperature fluctuations}
\author{J. {\L}uczka and B. Zaborek \\ 
Institute of Physics, University of Silesia,
    40-007 Katowice, Poland}
% P. Talkner\\
%Institut f\"{u}r Physik, Universit\"{a}t Augsburg,\\ 
%Universit\"{a}tsstr. 1, D-86135 Augsburg, Germany\\ }

%\date{}

\begin{abstract}
A diffusion process of a Brownian particle in a medium of temperature 
$T$ is re-considered. 
We assume that temperature of the medium fluctuates around its 
mean value. The velocity probability distribution is obtained. 
It is shown that  the stationary state is not a thermodynamic 
equilibrium state described by  the Maxwell distribution. Instead  
a nonequilibrium state is produced by temperature fluctuations. 

\end{abstract}

%\vspace{0.5cm}

\maketitle

%\newpage
% \baselineskip9mm

\section{Introduction}
\indent
Generalized statistics or 'superstatistics' occur in non-equilibrium systems 
as a result of parameter (temperature, friction, energy 
dissipation, pressure, chemical potential, etc.) fluctuations \cite{beck0}. 
An example of superstatistics is the 
Tsallis statistics in nonextensive  statistical mechanics \cite{tsal}.
One of a dynamical realization of this statistics has been  constructed 
by a Langevin 
equation for the Brownian particle \cite{wilk,beck} with the inverse 
temperature being a fluctuating parameter. In the paper we 
consider a more natural model with fluctuating temperature 
instead of its inverse. 
Fluctuations of temperature can play a significant role in many processes 
and phenomena. E.g., in astrophysics,  
the spectrum of temperature fluctuations of 
 the cosmic microwave background radiation can change our view on the universe 
at epochs from redshifts of the order of ten thousand to nearly the present 
and can provide important clues to inflationary models and the  dark 
matter-energy problem \cite{knel}. In plasma physics, 
an experimental evidence of substantial   temperature fluctuations 
has been found in mechanisms responsible for anomalous transport observed 
in tokamaks and stellarators \cite{hida}. The concept of temperature 
fluctuations is used in the theory of heavy ion collisions and 
multiparticle production \cite{stod}. In the Rayleigh-Benard convection, 
temperature fluctuations can be passively transported in the  turbulence 
regimes \cite{cioni}. Characteristics of temperature fluctuations 
in living tissue has been studied in  \cite{luba}. 
Below, we study the influence of temperature fluctuations 
on motion of the Brownian particle. 
As mentioned, the similar problem has been studied previously, 
mainly in the context of the 
Tsallis statistics \cite{wilk} with application to 
  velocity fluctuations in a turbulent flow \cite{beck}.
However, the studies have been limited to inverse temperature fluctuations and 
to the  'static' case when 
fluctuations are represented by a time-independent random variable. 
The problem is mathematically trivial in the sense that the 
probability density 
of velocity  can be calculated for an  arbitrary random variable 
modeling fluctuations.  
The more realistic model seems to be the  
'dynamic' model based on time-dependent noise for which 
temperature fluctuations 
are represented by a stationary stochastic process. Then, as is shown below, 
 the problem becomes non-trivial, even for the simplest model  of temperature 
fluctuations. 
 
Let us remind that in the classical theory of  diffusion, 
a position $x = x(t)$ of 
 a one-dimensional motion of a Brownian particle of mass $m$ moving 
in  an equilibrium  homogeneous medium of temperature $T$ 
is described by a Newton  equation  with a random force which,  
according to the fluctuation-dissipation theorem, has the form \cite{risk}
\begin{equation} \label{newton} 
 m\ddot{x}+\gamma \dot{x}=\sqrt{2\gamma kT}\:  \xi(t), 
\end{equation}
where $\gamma$ is the friction coefficient (given by e.g. the Stokes 
formula), $k$ is the Boltzmann constant  and $\xi(t)$ is a random force 
modeled by the Gaussian white noise,
\begin {eqnarray} \label{white}
 <\xi(t)>=0, \quad 
 <\xi(t)\xi(s)>=\delta(t-s).  
\end{eqnarray}
The velocity $v = \dot x$ is the  Ornstein-Uhlenbeck stochastic 
process and 
its probability density $P(v, t)$ obeys the Fokker-Planck equation
\begin {eqnarray} \label{F-P1}
{\frac{\partial P(v, t)}{\partial t}} = {\frac{\gamma}{m}} 
{\frac{\partial P(v, t)}{\partial v}} + {\frac{\gamma k T}{m^2}} 
\frac{\partial^2 P(v, t)}{\partial v^2}.
\end{eqnarray}
A general solution of this equation is given  by the expression 
\begin {equation} \label{gen}
P(v, t)=\int_{-\infty}^{\infty}p(v, t|v_{0}, 0)P(v_{0}, 0)\; dv_{0},
\end {equation}
where $P(v, 0)$ is an initial distribution and 
the transition probability distribution 
\begin {equation} \label{tra}
p(v, t|v_{0}, 0)  = \left[ 2 \pi \sigma ^2(t)\right]^{-1/2} 
\mbox{exp}\left\{-\left[v-v_0 \mbox{e}^{-\gamma t/m}\right]^2/2 \sigma ^2(t)
\right\}.
\end {equation}
The variance 
\begin {equation} \label{var}
 \sigma ^2(t) = {\frac{kT}{m}} \left(1-\mbox{e}^{-2\gamma t/m}\right).
\end {equation}
The stationary velocity distribution function $P_{M}(v)$ 
does not depend on the 
initial distribution $P(v, 0)$ and is the Maxwell distribution, 
\begin {equation} \label{max}
P_{M}(v)  = \sqrt{ {\frac{m}{2\pi kT}}} \; 
\mbox{exp}\left(-{\frac{mv^2}{2kT}}\right). 
\end {equation}
It means that the stationary state in the velocity space 
is a thermodynamic equilibrium  state.  

Now, let us consider the situation when the fluid is in a nonequilibrium 
steady state. What is a stationary state of the Brownian particle 
in this case? 
We should use a theoretical framework which enables to answer this 
question in a unified manner. For the moment, however, such a universal 
theory does not exist. First of all, we should specify a nonequilibrium state 
of the fluid: We assume that this state is not far from equilibrium and 
can be described similarly as an equilibrium state with the only 
exception that now temperature $T$ is time-dependent, $T=T(t)$. 
We follow and extend the proposal of Beck \cite{beck0,beck} from adiabatic 
to non-adiabatic temperature changes and apply Eq. (\ref{newton}) 
to this case.  
The problem is whether and when we can use Eq. (\ref{newton}) if $T=T(t)$,
cf. also polemics in \cite{oliv}.
Let us remember that when (\ref{newton}) is derived from the 
microscopic Hamiltonian model \cite{han0}, it is assumed that the 
fluid (medium) is in the thermodynamic equilibrium state of temperature $T$. 
If, however, the fluid is in the local-equilibrium steady state and  
can be characterized by the  time-dependent 
temperature $T=T(t)$,  then (\ref{newton}) cannot be rigorously justified. 
We are optimistic and believe that it can be used as a first approximation to 
an exact (but non-existing) theory.  
In section 2, we present a method how to treat such
 a system when temperature can fluctuate with $T(t)$  represented by 
a stochastic process 
and how to obtain the velocity distribution.  In section 3, 
we consider a curtailed  characteristic functional for which an evolution 
equation is determined by the infinitesimal generator of the 
stochastic process representing  temperature fluctuations. In section 4, 
we consider a case of dichotomic temperature fluctuations and solve the 
evolution equation for the curtailed characteristic functional. 
In section 5, we discuss properties of  the velocity  probability density. 
In section 6, we analyze statistical moments of the velocity. 

%%%%%%%%%%%%%%%%%%%%%%%%%%%%%%%%%%%%%%%%%%%%%%%%%%%%%%%%%%%%%%%%%%%%%%%%%%%%%

\section{Temperature fluctuations: Characteristic functional}

Now, let us assume that temperature of the fluid fluctuates around its 
mean value $T_0$, 
\begin{eqnarray} \label{fluT}
  T= T(t) = T_{0}+\eta(t). 
\end{eqnarray}
The zero-mean stationary stochastic process $\eta(t)$ describes 
temperature fluctuations and is independent of the stochastic process 
$\xi(t)$ describing thermal noise (interaction with surroundings). 
The formal  restriction  on this process follows from the condition 
$T(t)>0$ and  its phase space $Y$ is 
\begin{eqnarray} \label{spa}
 \eta(t) \in Y = (-T_0, \infty). 
\end{eqnarray}
The  velocity probability distribution $P(v, t)$ can be obtained from 
the relation (\ref{gen}), in which the initial transition probability density 
$p(v, t|v_{0}, 0)$  is expressed as  
\begin {equation} \label{p-C} 
p(v, t|v_{0}, 0) =\frac{1}{2\pi}\int_{-\infty}^{\infty}d\omega
e^{-i\omega v} C_{v}(\omega, t; v_0), 
\end {equation}
where the conditional characteristic 
function $C_v(\omega, t; v_0)$ of the velocity is defined by 
\begin {equation} \label{Cv}
C_v(\omega, t; v_0) = \left< \mbox{e}^{i\omega v(t)}\right>.  
\end {equation}
The velocity $v(t)$ is a solution of Eq. (\ref{newton}) with 
time-dependent temperature $T=T(t)$, namely,  
\begin {eqnarray} \label{solne}
v(t)=v_{0}\exp\left(-\gamma
t/m\right)+\sqrt{2k\gamma/m^2} \: \int_{0}^{t}ds
\exp\left[-\gamma(t-s)/m\right]\sqrt{T(s)}\:
\xi(s),\nonumber\\
\ \  
\end {eqnarray}
where $v_0$ is the initial velocity of the Brownian particle. 
Inserting the above equation into (\ref{Cv}) yields 
\begin{eqnarray} \label{cha2}
C_{v}(\omega, t; v_0) = 
 \mbox{exp}\left(i\omega v_{0}e^{-\gamma t/m}\right)\: C(\omega, t),
\end{eqnarray}
where
\begin{eqnarray}
C(\omega,t)=\left< \exp \left[i\omega \sqrt{2k\gamma/m^2} 
\int_{0}^{t}ds e^
{-\gamma(t-s)/m}
\sqrt{T(s)} \: \xi(s)\right]\right>_{\xi, \eta}.
\label{14}
\end{eqnarray}
The subscripts  $\xi$  and $\eta$ denote average over all realizations of
thermal noise $\xi(t)$ and temperature fluctuations  $\eta(t)$, 
respectively. The averaging over the Gaussian white noise $\xi(t)$ 
can be performed leading to the expression   
\begin{eqnarray} \label{after}
C(\omega,t)= \exp{\left[-(kT_0/2m) \omega ^2 
\left(1-\mbox{e}^{-2\gamma t/m}\right)\right]} \Phi_{\eta}(\omega, t),
\end{eqnarray}
where 
\begin{eqnarray} \label{Phi}
\Phi_{\eta}(\omega, t)=
\left<\exp\left[- (\gamma  k/m^2) \omega ^2  
\mbox{e}^{-2\gamma t/m}
\int_{0}^{t}ds \: \mbox{e}^{2 \gamma s/m} \eta(s)\right]\right>_{\eta} 
\end{eqnarray}
is the characteristic functional of the stochastic process $\eta(t)$. 
In this approach, the velocity probability of the Brownian particle 
is determined by the characteristic functional of temperature fluctuations. 
The explicit form of this functional will be obtained by the method of the 
so-called 'curtailed' characteristic functional. 

%%%%%%%%%%%%%%%%%%%%%%%%%%%%%%%%%%%%%%%%%%%%%%%%%%%%%%%%%%%%%%%%%%%%%%%%%%%
\section{Curtailed characteristic functional} 

In order to calculate the functional (\ref{Phi}) we proceed in the 
following way \cite{lucz}. 
For fixed time $t=\tilde{t}$ we define \cite{sib}  
\begin{eqnarray} \label{Omega}
\Omega  = (\gamma k/m^2) \omega ^2 \mbox{e}^{-2\gamma \tilde{t}/m }.
\end{eqnarray}
Let us introduce the auxiliary functional 
\begin{eqnarray} \label{Psi}
 \Psi [\eta; \Omega, \tilde{t}] = 
\left<\exp\left[- \Omega 
\int_{0}^{\tilde{t}}ds \: \mbox{e}^{2 \gamma s/m} 
\eta(s)\right]\right>_{\eta}. 
\end{eqnarray}
Then the relation 
\begin{eqnarray} \label{rela2}
\Phi_{\eta}(\omega, t) = \Psi[\eta; \Omega, \tilde{t}=t]
\end{eqnarray}
holds. \\
The curtailed characteristic functional corresponding to (\ref{Psi}) 
is defined as \cite{kamp}
\begin {eqnarray} \label{curt}
V(y, t)=\left<\delta(\eta(t), y) \exp\left[-\Omega\int_{0}^{t} ds
e^{2 \gamma s/m}\eta(s)\right]\right>_{\eta},  
\end{eqnarray}
where  $y\in Y$ takes values from the phase space  of 
the stochastic process $\eta(t)$ and 
$\delta(\eta(t), y)$ is the Kronecker delta when  $\eta(t)$ is a  discrete 
process or the Dirac delta  for the  continuous 
stochastic processes $\eta(t)$.  
The relation between these two functions is the following 
\begin{eqnarray} \label{beet}
\Phi_{\eta}(\omega, t) = \int_{Y} V(y, t) dy,
\end{eqnarray}
where the integration for the continuous (or summation for discrete ) 
process is over the phase space $Y$.  
We introduce  curtailed characteristic  functional because for it, 
in contrary to (\ref{Phi}), 
an evolution equation is known. In the abbreviated notation, it has the form 
\cite{kamp} 
\begin{eqnarray} \label{evol1}
\frac{\partial}{\partial t}V(y, t)={\hat L} V(y, t)-\Omega
e^{2 \gamma t/m}\: y V(y,  t),
\end{eqnarray}
where ${\hat L}$ is an infinitesimal generator  (a forward operator) 
of the stochastic 
process $\eta(t)$. If $\eta(t)$ is determined  by an Ito stochastic equation, 
the  infinitesimal generator is a differential operator which occurs in 
the {\it forward Kolmogorov equation} (i.e. in the Fokker-Planck equation).
Now, the problem reduces to solving the evolution equation (\ref{evol1}) 
which, in dependence of $\eta(t)$, can be a single  or a set of  
differential or  integro-differential equations. Below, we present an 
example which can be solved exactly. 
%
%\begin{eqnarray}
% {\hat L} = \left(\begin{array}{cc}
% -\lambda & \mu \\
%  \lambda &-\mu
%\end{array}\right)
%\end{eqnarray}
% 
%%%%%%%%%%%%%%%%%%%%%%%%%%%%%%%%%%%%%%%%%%%%%%%%%%%%%%%%%%%%%%%%%%%%%%%%%%%%%%%

\section{Dichotomic fluctuations}

Here, we consider a caricature of temperature fluctuations, i.e. a discrete, 
two-state model \cite{beck0}. 
An extension  to a many-state or  continuous model of fluctuations 
is in principle  possible \cite{lucz0}. However, physics should be similar but 
mathematics would be  much more complicated because of difficulties 
in solving the evolution equation (\ref{evol1}).  
So, we represent temperature fluctuations by dichotomic noise \cite{hang1}
\begin{eqnarray} \label{dich}
\eta(t)= \{-a , b\}, \quad 0 < a < T_0, \quad b > 0.
\end{eqnarray}
Transition probabilities per unit time from one state to the other are
given by the relations
\begin{eqnarray} \label{jum}
 Pr(-a\rightarrow b)=\mu = 1/\tau_a,
\quad Pr(b\rightarrow -a)=\lambda = 1/\tau_b, 
\end{eqnarray}
where $\tau _a$ and $\tau _b$ are mean waiting times in states $-a$ and $b$, 
respectively. We assume that 
\begin{equation} \label{zero}
b \mu= a \lambda. 
\end{equation}
Then the process is stationary and the probabilities 
\begin{eqnarray} \label{sta}
Pr(\eta(t)=-a)=\frac{\lambda}{\mu+\lambda} = \frac{b}{a+b}, \quad 
Pr(\eta(t)=b)=\frac{\mu}{\mu+\lambda} =\frac{a}{a+b}.
\end{eqnarray}
The  first two moments read
\begin{eqnarray} \label{mom}
\langle \eta (t) \rangle = 0, \quad  
\langle \eta (t) \eta(s) \rangle = a b \:\mbox{exp} 
\left(-|t-s|/\tau_c \right), 
\end{eqnarray}
where the correlation time $\tau_c$ is given by the formula 
$1/\tau_c = \mu + \lambda$.  

The relation (\ref{beet}) takes the form 
\begin{eqnarray} \label{beetd}
\Phi_{\eta}(\omega, t) = V(-a, t) + V(b, t)
\end{eqnarray}
and  the explicit form of (\ref{evol1}) reads 
\begin{eqnarray}\label{e2}
\frac{\partial}{\partial t}V (-a,t) =-\mu V (-a,t)+\lambda V
(b,t)+\Omega e^{2\gamma t/m} a  V (-a,t),
\end{eqnarray}
\begin{eqnarray} \label{e1}
\frac{\partial}{\partial t}V (b,t) =\mu V (-a,t)-\lambda V
(b,t)-\Omega e^{2\gamma t/m} b  V (b,t).
\end{eqnarray}
The initial conditions follow from (\ref{curt}) and read (cf. (\ref{sta}))
\begin{eqnarray}
 V (-a,0)= \left<\delta(\eta(t), -a)\right>= \frac{\lambda}{\mu+\lambda},
\end{eqnarray}
\begin{eqnarray}
 V (b,0)= \left<\delta(\eta(t), b)\right>= \frac{\mu}{\mu+\lambda}.
\end{eqnarray}
Let us define a new time  variable 
\begin{eqnarray} \label{tau}
\tau = \tau(t)=\Omega e^{2\gamma t/m}
\end{eqnarray}
and two new functions $\tilde{V}(-a,\tau)$ and $\tilde{V}(b,\tau)$  
via the relations 
 \begin{eqnarray}\label{a2}
 V (-a,t)=\tilde{V}
(-a,\tau(t)),
\end{eqnarray}
\begin{eqnarray}\label{a1}
V (b,t)=\tilde{V} (b,\tau(t)).
 \end{eqnarray}
Then Eqs (\ref{e1}) and (\ref{e2}) can be transformed to the form 
\begin{eqnarray}\label{b2}
\frac{2\gamma \tau}{m} \frac{\partial}{\partial \tau}\tilde{V}
(-a,\tau)=-\mu \tilde{ V} (-a,\tau)+\lambda \tilde{V}
(b,\tau)+\tau a  \tilde{V} (-a,\tau),
 \end{eqnarray}
\begin{eqnarray}\label{b1}
\frac{2\gamma \tau}{m} \frac{\partial}{\partial \tau}\tilde{V} (b,\tau)=\mu
\tilde{ V} (-a,\tau)-\lambda \tilde{V} (b,\tau)-\tau b  \tilde{V}
(b,\tau)
 \end{eqnarray}
with the initial conditions at $\tau(t=0)=\Omega$, 
\begin{eqnarray}
\tilde{ V} (-a,\Omega)= \frac{\lambda}{\mu+\lambda},
\end{eqnarray}
\begin{eqnarray}
\tilde{ V} (b,\Omega)= \frac{\mu}{\mu+\lambda}.
\end{eqnarray}
We define two new functions in the following way 
\begin{eqnarray} \label{F}
F(\tau)=\tilde{V} (-a,\tau) + \tilde{V} (b,\tau),
\end{eqnarray}
\begin{eqnarray} \label{G}
G(\tau)= b \tilde{V} (b,\tau)- a \tilde{V} (-a,\tau).
\end{eqnarray}
Then from Eqs (\ref{b2}) and (\ref{b1}) one gets 
\begin{eqnarray} \label{dotF}
\frac{2\gamma}{m}  \dot{F}(\tau)= - G(\tau),
\end{eqnarray}
\begin{eqnarray} \label{dotG}
\frac{2\gamma \tau}{m}  \dot{G}(\tau)+(\mu+\lambda-\tau(a-b))G(\tau)+\tau a b
F(\tau)=0, 
\end{eqnarray}
where the dot denotes a derivative with respect to the argument. 
The initial conditions follow from (\ref{F})-(\ref{dotG}) and take the form 
\begin{eqnarray} \label{iniFG}
F(\Omega) = 1, \quad G(\Omega) = 0.
\end{eqnarray}
The function $F(\tau)$ is crucial because the characteristic functional 
(\ref{Phi}) is related to it in a simple way. Indeed, 
\begin{eqnarray} \label{rela}
\Phi_{\eta}(\omega, t) = F(\tau)  \quad \mbox{for} \quad  
\tau = \Omega e^{2\gamma t/m} \quad \mbox{and} \quad
\Omega  = (\gamma k/m^2) \omega ^2 \mbox{e}^{-2\gamma t/m }.
\end{eqnarray}
From the above system of two coupled differential equations (\ref{dotF}) and 
(\ref{dotG}), we obtain 
a closed differential equation for the function $F(\tau)$ only. 
It has the form 
\begin{eqnarray} \label{dotdotF}
\tau
\ddot{F}(\tau)+\frac{m}{2\gamma}\left[\mu+\lambda + \tau (b-a)\right]
\dot{F}(\tau)-
\frac{m^2 ab\tau }{4\gamma^2}F(\tau)=0
\end{eqnarray}
with the initial conditions 
\begin{eqnarray} \label{iniF}
F(\Omega) = 1, \quad \dot{F}(\Omega) = 0.
\end{eqnarray}
It belongs to a class of hypergeometric equations. Its solution is 
the function \cite{kamke} 
\begin{eqnarray}\label{rozwh}
F(\tau)=\mbox{e}^{-m b \tau /2 \gamma}\left\{
C_{1}(\Omega)\;\Phi[\alpha,\beta,\chi(\tau)]+
C_{2}(\Omega)\;\Psi[\alpha,\beta,\chi(\tau)]\right\},
\end{eqnarray}
where $\Phi$ and $\Psi$ stand for the confluent hypergeometric 
Kummer and Tricomi functions,  respectively \cite{ober}. The parameters
\begin{eqnarray}
\alpha=\frac{m b}{2\gamma}\frac{\mu+\lambda}{a+b}=\frac{m b}{2\gamma
\tau_{c}(a+b)},
\end{eqnarray}
\begin{eqnarray}
\beta=\frac{m(\mu+\lambda)}{2\gamma}=\frac{m}{2\gamma\tau_{c}}
\end{eqnarray}
and 
\begin{eqnarray} \label{defl}
 \chi(\tau)= \frac{m (a+b)}{2\gamma} \; \tau.
\end{eqnarray}
The constants  $C_{1}(\Omega)$ and  $C_{2}(\Omega)$ are 
determined from the  conditions (\ref{iniF}) and read 
\begin{eqnarray}\label{std}
 C_{1}(\Omega)&=&\frac{b \Gamma(\alpha)}{(a+b) \Gamma(\beta)}\:
\chi(\Omega)^{\beta}\:\mbox{e}^{- m a \Omega/2\gamma}\nonumber\\
&\times&\left(\Psi[\alpha,\beta,\chi(\Omega)]+ 
\frac{m(\mu + \lambda)}{2\gamma}\; 
\Psi[\alpha+1,\beta+1,\chi(\Omega)]\right)
\end{eqnarray}
and 
\begin{eqnarray}\label{stg}
C_{2}(\Omega)&=&    \frac{b \Gamma(\alpha)}{(a+b) \Gamma(\beta)}\:
\chi(\Omega)^{\beta}\:\mbox{e}^{- m a \Omega/2\gamma}\nonumber\\
&\times& \left(\Phi[\alpha+1,\beta+1,\chi(\Omega)]-
\Phi[\alpha,\beta,\chi(\Omega)]\right),
\end{eqnarray}
where $\Gamma(z)$ is the Euler Gamma function.  
In this way we found the  function $F(\tau)$  and via the expressions in 
({\ref{rela}) we can find the characteristic functional 
$\Phi_{\eta}(\omega, t)$.

%%%%%%%%%%%%%%%%%%%%%%%%%%%%%%%%%%%%%%%%%%%%%%%%%%%%%%%%%%%%%%%%%%%%%%%%%%%%%%%

\section{Probability distribution} 

The probability distribution is obtained from Eqs (\ref{p-C})-(\ref{Phi}) 
and the relations ({\ref{rela}). It has the form 
\begin{eqnarray}\label{genrozw}
p(v,t|v_{0},0)&=&\frac{1}{2\pi}\int_{\infty}^{\infty}d\omega\;\mbox{e}^{i\omega
v }\nonumber\\
&\times&\exp\left[i\omega v_{0}\mbox{e}^{-\gamma t/m }-(kT_0/2 m) \omega ^2 
\left(1-\mbox{e}^{-2\gamma t/m}\right)\right] \Phi_{\eta}(\omega, t).\; 
\end{eqnarray}
The  explicit form  of the characteristic functional is  
\begin{eqnarray}\label{F(tau)pelne1}
\Phi_{\eta}(\omega, t)= \frac{b \Gamma(\alpha)}{(a+b) \Gamma(\beta)}\:
\mbox{e}^{-bk\omega^{2}/2m}  
\left(A\omega^{2}\mbox{e}^{-2\gamma t/m }\right)^{\beta} 
\exp\left[-(ak\omega^2/m)\mbox{e}^{-2\gamma\;t/m}\right]
\nonumber\\
\times \Bigg\{\Phi\left[\alpha, \beta,  A\omega^{2}\right] 
\Bigg(\Psi\left[\alpha, \beta, A\omega^{2}\mbox{e}^{-2\gamma t/m }\right]
+\frac{m}{2\gamma \tau_c} 
\Psi\left[\alpha+1, \beta+1, A\omega^{2}\mbox{e}^{-2\gamma t/m }
\right]\Bigg)\nonumber\\
+\Psi\left[\alpha, \beta, A\omega^{2}\right]
\Bigg(\Phi\left[\alpha + 1, \beta + 1, 
A\omega^{2}\mbox{e}^{-2\gamma t/m}\right]
-\Phi\left[\alpha, \beta, A\omega^{2}\mbox{e}^{-2\gamma t/m
}\right]\Bigg)\Bigg\},\nonumber\\
\ \ \ \ \:\;\
\end{eqnarray}
where the constant $A=k(a+b)/2m$. 
By use of (\ref{gen}), 
the distribution $p(v,t|v_{0},0)$ allows to determine evolution
of the one-dimen\-sio\-nal velocity probability density $P(v,t)$ for an 
arbitrary initial state defined by the distribution  $P(v,0)$ and analyze 
 relaxation of the system to the stationary state. 

%%%%%%%%%%%%%%%%%%%%%%%%%%%%%%%%%%%%%%%%%%%%%%%%%%%%%%%%%%%%%%%%%%%%%%%%%%%%%%%%%%

\subsection{ Stationary distribution}

The stationary velocity distribution function $P_{st}(v)$ 
does not depend on the initial distribution $P(v, 0)$.  
 It is obtained from (\ref{gen}) and (\ref{genrozw}) 
performing the long time limit, 
 $t\rightarrow\infty$. We use the relations \cite{ober}
\begin{eqnarray}\label{limkumer}
\lim_{z \rightarrow 0}\;\Phi[\alpha,\beta,z]=1
\end{eqnarray}
and 
\begin{eqnarray}\label{limtricomi}
\Psi[\alpha,\beta,z]=\frac{\Gamma(1-\beta)}{\Gamma(\alpha-\beta+1)}
+\frac{\Gamma(\beta-1)}{\Gamma(\alpha)}\;z^{1-\beta}, 
\end{eqnarray}
which represents the leading terms for small 
$z = A\omega^{2}\mbox{e}^{-2\gamma t/m}  \ll 1$ when $t\to\infty$.    
Then the stationary distribution takes the form
\begin{eqnarray}\label{st}
P_{st}(v)=\frac{1}{\pi}\int_{0}^{\infty}d\omega \cos(\omega v)
\;\mbox{e}^{-(T_{0}+b)k \omega^{2}/2
m}\;\Phi\left[\frac{b\;m}{2\gamma\tau_{c}(a+b)},\frac{\;m}{2\gamma\tau_{c}},
\frac{a+b}{2\;m}\;k\;\omega^{2}\right].\nonumber\\
\ \ \
\end{eqnarray}
It is not a Maxwell distribution and we can conclude that the stationary state 
is not an equilibrium state: It is a nonequilibrium state. 
In the case of absent of temperature fluctuations, i.e. when $a=0$ and 
next $b=0$, then 
\begin{eqnarray}\label{Phi0}
\Phi\left[\frac{m}{2\;\tau_{c}\gamma},\frac{\;m}{2\gamma
\tau_{c}},0\right]=1
\end{eqnarray}
and the Maxwell distribution (\ref{max})   is obtained for  $T=T_{0}$.

%%%%%%%%%%%%%%%%%%%%%%%%%%%%%%%%%%%%%%%%%%%%%%%%%%%%%%%%%%%%%%%%%%%%%%%%%%%%%%

\subsection{ Limiting cases of short and long correlation time}

The correlation time $\tau_c$ of temperature fluctuations is defined below Eq. 
(\ref{mom}). For very fast fluctuations when the correlation time is short, 
\begin{eqnarray}\label{kumertaukrotkie}
\lim\limits_{\tau_{c}\rightarrow 0} 
\Phi\left[\frac{b\;m}{2\gamma\tau_{c}(a+b)},\frac{\;m}{2\gamma\tau_{c}},\frac{a+b}{2\;m}\;k\;\omega^{2}\right]=
\;\mbox{e}^{b k\;\omega^{2}/2 m}
\end{eqnarray}
and (\ref{st}) reduces to the Maxwell distribution (\ref{max}) 
with temperature 
$T=T_0$.   
The short correlation time limit can be achieved when (cf. (25)): \\ 
(i) $\mu\rightarrow\infty$ and  $b\rightarrow 0$ but   
$b\mu=a\lambda=const.$ \\
(ii) $\;\lambda\rightarrow\infty$ and $a \rightarrow 0$ but 
$a\lambda=b\mu=const.$\\
(iii) $b\to \infty$ and $\lambda \to\infty$ but $b/\lambda = const. $
(the latter corresponds to the Poisson 
white shot noise)\\
(iv) $\mu\rightarrow\infty$ and $\lambda\rightarrow\infty$ but 
$\mu/\lambda = const.$ \\ 
In these limits, the system is not able to react to very fast fluctuations and 
effectively it feels the mean temperature $T=T_0$. 

The opposite limit is the adiabatic limit when fluctuations are slow and 
the  correlation time is very long,  $\tau_{c}\rightarrow\infty$. The Kummer 
function takes the form 
\begin{eqnarray}\label{kumertaudlugie}
\lim\limits_{\tau_{c}\rightarrow \infty} 
\Phi\left[\frac{b\;m}{2\gamma\tau_{c}(a+b)},\frac{\;m}{2\gamma\tau_{c}},\frac{a+b}{2\;m}\;k\;\omega^{2}\right]=
\frac{a}{a+b}+\frac{b}{a+b}\;\mbox{e}^{(a+b)k\omega^{2}/2m}
\end{eqnarray}
and Eq. (\ref{st}) reduces to the function
\begin{eqnarray}\label{prtauduze}
P_{st}(v) =  \frac{b}{a+b}\sqrt{\frac{m}{2\pi
k(T_{0}-a)}}\;\exp\left[-\;\frac{m v^{2}}{2k(T_{0}-a)}\right]\nonumber\\
+\frac{a}{a+b}\sqrt{\frac{m}{2\pi
k(T_{0}+b)}}\;\exp\left[-\;\frac{m v^{2}}{2k(T_{0}+b)}\right].
\end{eqnarray}
It is a linear combinations of two Maxwellian distributions 
for two temperatures  $T_{0}-a$ i $T_{0}+b$ and with the weights 
given by Eqs (\ref{sta}).  
The long correlation time limit can be achieved when 
$\mu,\;\lambda$ $\rightarrow0$ and the mean residence times in the states 
$-a$ and  $b$ tend to infinity, $\tau_{a},\;\tau_{b}\rightarrow\infty$.
The adiabatic limit for other models of inverse temperature 
fluctuations has been 
considered in \cite{beck0}.

\begin{figure}[h]
%noindent
\begin{center}
\includegraphics[angle=0,scale=0.75]{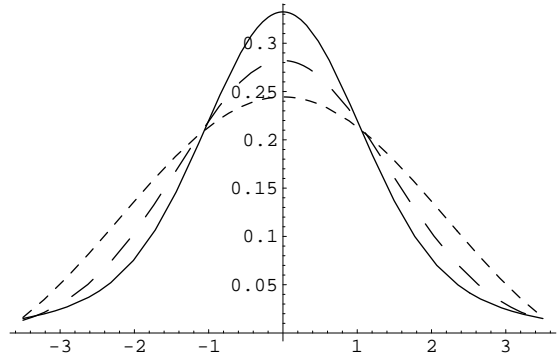}
\caption{\footnotesize {The stationary velocity distribution $P_{st}(v)$
for  three values of the  correlation time $\tau_c$ 
of temperature fluctuations: $\tau_c=0$ (dotted line, the  Maxwell 
distribution),  $\tau_c =5$ (dashed line) and $\tau_c \to\infty$ 
(solid line).}}
\end{center}
\end{figure}

%%%%%%%%%%%%%%%%%%%%%%%%%%%%%%%%%%%%%%%%%%%%%%%%%%%%%%%%%%%%%%%%%%%%%%%%%%%%%%%%%

\section{Discussion}

Applying a method of the curtailed characteristic functional we 
obtained a time-dependent probability distribution of  velocity of 
the Brownian particle moving in medium in which temperature fluctuates. 
We considered temperature fluctuations to be as simple as possible, i.e.,  
 a two-state   stationary Markovian  process $\eta(t)$. Nevertheless, the  
problem is formulated for an arbitrary Markovian stochastic process 
$\eta(t)$ because what we need is the infinitesimal generator 
${\hat L}$ of the process $\eta(t)$, see Eq. (\ref{evol1}).  
We note that the correlation function of the force 
$F(t)= \sqrt{2\gamma kT(t)}\:  \xi(t)$ in Eq. (\ref{newton}) has the form 
\begin {eqnarray} \label{F(t)}
 <F(t)F(s)>= 2\gamma kT_0 \delta(t-s), 
\end{eqnarray}
independently of statistics of fluctuations $\eta(t)$ and has the same 
 form as  in the case without temperature fluctuations. 
It resembles the dissipation-fluctuation relation.   
However, 
we showed that the stationary state is a nonequilibrium state. We 
can  ask how far the system is from equilibrium. To this aim, 
let us analyze statistical moments of the velocity.  
From Eq. (\ref{st}) it follows that the stationary characteristic function 
$C_{v}(\omega)$ of the velocity is 
\begin{eqnarray}\label{Cst}
C_{v}(\omega)= \mbox{e}^{-(T_{0}+b)k \omega^{2}/2
m}\;\Phi\left[\frac{b\;m}{2\gamma\tau_{c}(a+b)},
\frac{\;m}{2\gamma\tau_{c}},\frac{a+b}{2\;m}\;k\;\omega^{2}\right].
\end{eqnarray} 
The statistical moments $\langle v^n \rangle$, $(n=1, 2, 3, ...)$ can 
be obtained from the relation 
$<\langle v^n \rangle>= i^n d^nC_{v}(\omega)/d\omega^n\vert_{\omega = 0}$. 
The odd order moments are equal to zero. The second moment 
\begin{eqnarray}\label{m2}
\langle v^2 \rangle = kT_0/2.
\end{eqnarray} 
It does not depend on the statistics of temperature fluctuations and 
is the same as for the Maxwell distribution! The forth order moment measure 
a deviation from the Maxwell distribution. We use it to calculate 
the kurtosis,  
\begin{eqnarray}\label{m4}
Kurt(v) = \frac{\langle v^4 \rangle}{3 \langle v^2 \rangle ^2} - 1 = 
\frac{2 a b}{T_0^2 (2+\tau_d/\tau_c)},  
\end{eqnarray} 
where $\tau_d = m/\gamma$ is the relaxation time of the velocity 
in the deterministic case (cf. Eq. (\ref{solne}) when $\xi(t) = 0)$ 
and $\tau_c$ is the correlation time of fluctuations (see Eq. (\ref{mom})).  
For the equilibrium state, i.e. for the Maxwell distribution, 
the kurtosis is zero. In the case considered, 
the kurtosis is always positive and it means that $P_{st}(v)$ is more peaked 
than the Maxwell distribution. 
It is an increasing function of the variance $<\eta^2(t)> = ab$ 
and the correlation time $\tau_c$  of temperature fluctuations. 
As a function of the correlation time, 
it grows from zero for $\tau_c =0$ to the maximal value $ab/T_0^2$ 
when $\tau_c \to \infty$. Generally, all even order moments 
are greater than for the Maxwell distribution. 

One can determine the moments for the position of the Brownian 
particle. E.g., for  long times, $t\gg \tau_d$, 
the mean squared displacement 
\begin{eqnarray}\label{M2}
\langle x^2(t) \rangle \sim 2Dt, 
\end{eqnarray} 
where the diffusion coefficient $D=kT_0/\gamma$ is the same  as in 
the case without 
temperature fluctuations. It means that for long time, 
the process in the position space 
is the standard normal diffusion with the same diffusion constant.   
We can conclude that the  first two moments of position and of velocity 
are the same in both cases: without and with temperature fluctuations. 
So, when we  measure only first two moments, we can not distinguish 
these two states. We emphasize that it does not depend on the model of 
temperature fluctuations.  \\

\noindent The work partially supported by the European Science 
Foundation (the Program 
{\it Stochastic Dynamics: Fundamentals and Applications}).

\end{document}